# Behavior-Based Detection of GPU Cryptojacking

Dmitry Tanana
*Laboratory of Combinatorial Algebra*
Ural Federal University
Yekaterinburg, Russia
ddtanana@urfu.ru

*Abstract*—With the surge in blockchain-based cryptocurrencies, illegal mining for cryptocurrency has become a popular cyberthreat. Host-based cryptojacking, where malicious actors exploit victims' systems to mine cryptocurrency without their knowledge, is on the rise. Regular cryptojacking is relatively well-known and well-studied threat, however, recently attackers started switching to GPU cryptojacking, which promises greater profits due to high GPU hash rates and lower detection chance. Additionally, GPU cryptojackers can easily propagate using, for example, modified graphic card drivers. This article considers question of GPU cryptojacking detection. First, we discuss brief history and definition of GPU cryptojacking as well as previous attempts to design a detection technique for such threats. We also propose complex exposure mechanism based on GPU load by an application and graphic card RAM consumption, which can be used to detect both browser-based and host-based cryptojacking samples. Then we design a prototype decision tree detection program based on our technique. It was tested in a controlled virtual machine environment with 80% successful detection rate against selected set of GPU cryptojacking samples and 20% false positive rate against selected number of legitimate GPU-heavy applications.

## I. Introduction

In 2024 global cryptocurrency market capitalization has reached more than 2.2 trillion dollars [1]. Mining of cryptocurrencies, or generation process for new cryptocoins usually requires enormous amount of processing power, with largest of all cryptonetworks – Bitcoin – generating around 600 million terahashes per second in May 2024 [2], while consuming more than 100 TWh in 2023 [3]. The popularity of cryptocurrencies makes them a valued target for all sorts of cybercriminals. Due to their relative untraceability, cryptocoins are extensively used in shadow economy – from drug deals to ransom demands [4]. But those qualities also attract malicious actors to make unauthorized use of victims' machines in order to mine cryptocurrencies. Such attacks are called cryptojacking or malicious mining.

Malicious mining attacks can be divided into two main categories [5]:

1. Executable-type cryptojackers – usually a converted legitimate miner with customizable configuration parameters. It is installed on victims' system via spam, exploit kits or other malware.
2. Browser-based cryptojackers – occurs in users' web browser when they visit a web page with an active mining script. Under most jurisdictions that is considered legal if website obtains user's consent.

Browser-based mining was first introduced to wide public by Coinhive project to mine Monero cryptocurrency. The authors' original idea was to provide website owners with alternative to advertisement source of income, however it quickly become very popular with cybercriminals. After Coinhive was shut down in 2019, the number of browser-based cryptojacking attacks has decreased considerably until cybercriminals switched to other mining services [6].

Nowadays, cryptojacking attacks are especially prominent in the financial sector. According to more recent report by SonicWall, the number of cryptojacking attacks in the finance companies has increased by 269% since the beginning of 2022. Overall, the number of cryptojacking incidents increased by 30% in the first half of 2023. Hovever, the number of attacks on the financial industry is five times higher than in the second half of 2022 [7].

Another recently emerged threat in cryptojacking, which will be the focus of this study is GPU cryptojacking. For the most of its history, cryptojacking used CPU resources because they're much easier to access from browser [8]. However, GPUs are a more lucrative target for cybercriminals due to their higher hash-rates and larger pool of cryptocurrencies to mine. Additionally, previous studies showed that GPU cryptojackers can easily propagate through modified software like Kubernetes clusters and NVIDIA drivers [9, 10].

While GPU cryptojacking doesn't directly harm infected users, victims still incur losses as their hardware degrades and their processing power is diverted toward cybercriminals' purposes rather than the legitimate work those processing units were supposed to do. For businesses, especially in technological sector, GPU cryptojacking poses an even greater threat. If a cluster of computers becomes infected, the business may suffer significant financial losses due to increased electricity consumption and the need to replace overworked processing units. Those problems are amplified by current surge of interest in AI computations, which have numerous applications in many spheres and also done primarily on GPUs, therefore increasing target accessibility for GPU cryptojacking.

So, the early detection of GPU cryptojacking is crucial to minimize losses for victims. In this paper, we propose a new method for detecting executable-type GPU cryptojackers based on the average quadratic deviation of GPU processing power usage and a number of other metrics. Additionally, we present a prototype tool designed to identify potential GPU cryptojackers on Windows OS.

## II. Previous works

To the best of our knowledge, there're only two papers which consider GPU cryptojacking detection. The first one is "Overcoming the Pitfalls of HPC-based Cryptojacking Detection in Presence of GPUs" by C. Pott et al. [11]. The second one – "MagTracer: Detecting GPU Cryptojacking Attacks via Magnetic Leakage Signals" by R. Xiao et al. [12].

In MagTracer, the authors propose to detect magnetic signals emanating from GPUs when they're under mining

This work was supported by the Ministry of Science and Higher Education of the Russian Federation, project no. FEUZ-2023-022.

load, those signals are produced by all cryptomining algorithms – they're all compute-intensive and memory bounded. In their experiments, R. Xiao and his colleagues observed that during the mining process, GPUs produce high magnitude currents, up to 30A, which result in strong magnetic fields, on the order of $100\mu T$ close to GPU. They then assemble a prototype sensor system to detect those fields and test it on 14 different GPU models, with true positive rate of more than 98% and false positive rate of less than 0,7%. In our opinion this is a very creative solution, but its main advantage – external hardware sensor is also its main disadvantage. The remote cybercriminal cannot tamper with external sensor, so even if attacker obtains full control of victims' system, he'll still be detected. At the same time, external hardware sensor might be hard to deploy on remote server or, if there's need to monitor business' network of computers, then multiple sensors connected into their own IoT network have to be installed, which would require significant investments [11].

C. Pott's team uses previous cryptojacking research to demonstrate that Hardware Performance Counters (HPCs) can be used to detect CPU cryptojacking attacks with 96% accuracy. Then they demonstrate that CPU HPCs are not sufficient to detect GPU Cryptojackers. Finally, the authors improve previous detection approach, adding GPU performance metrics to achieve 99,86% detection rate for GPU cryptojackers, while having very low resource consumption for the monitoring system, which allows for continuous surveillance of live systems. They've collected the following GPU metrics with NVIDIA GPU monitoring tool: fan speed, temperature, power usage, memory utilization and GPU utilization and then used them for training of Machine Learning Classifiers. In our opinion, some of those metrics are unreliable and/or redundant. For example, fan speed and temperature are closely connected, while maximum and working temperature also depends on room temperature, GPU model, hardware setup and even thermal paste used [12].

Finally, we would also be leveraging methods used in our previous work about CPU cryptojacking detection – Advanced behavior-based technique for cryptojacking malware detection [13].

## III. DETECTION ALGORITHM

### A. Instruments

To ensure successful operation, the cryptojacking detection program should realize these steps:

1. Collect data on the analyzed machine
2. Identify signs of infection within that data
3. Conduct further analysis to determine if there is potential cryptojacking attack running on machine in question

To implement those steps, we must first identify necessary signs of infection based on cryptojacker behavior. Additionally, we need an algorithm to make the final decision – whether the machine is infected or not.

For gathering data on GPU metrics, we will be using NVIDIA System Management Interface [14], which is based on top of the NVIDIA Management Library. It allows us to monitor all important GPU metrics for a selected process, which is very important if multiple GPU-dependent applications are running on same machine.

Furthermore, since NVIDIA System Management Interface operates at a low level, it can bypass certain stealth techniques employed by cryptojackers. For instance, some cryptojacking samples go idle when users launch Task Manager, therefore evading behavioral detection by ordinary users.

To identify indicators of cryptojacking activity in Windows we conducted an experiment in a controlled machine environment. The host machine had Windows 11 with Intel quad-core CPU, 16 GB RAM and NVIDIA GTX 3060 GPU with 12 GB memory. Unfortunately, we were unable to utilize virtualization technology, since it offers poor GPU performance, therefore the testing was not done on live cryptojacking specimens, but rather legitimate mining applications which are popular with cryptojacking cybercriminals. In that environment we've analyzed 3 basic XMRig configurations – XMRig being responsible for vast majority of GPU cryptojacking attacks [7]. Also, we've analyzed few legitimate GPU-heavy applications: Blender – a 3D creation suite, FurMark – GPU benchmark and Elden Ring – modern graphics videogame.

### B. Infection indicators

We have initiated the identification of cryptojacking infection indicators by analyzing computer resource consumption.

Our experiments revealed that GPU mining, unsurprisingly, heavily relies on graphic processor usage, with average processor load of around 92%. Unfortunately, some most legitimate GPU-intensive applications also utilize graphic processor extensively, rendering detection of GPU cryptojacking malware based solely on GPU usage impossible.

The second indicator we have considered was the amount of RAM consumed by process. Our findings indicate that processes associated with cryptojacking activity consume 3.7 GB of RAM on average, with 3.1 GB being lowest and 4.2 being highest. It should be moted, that mining processes take a nearly constant amount of RAM, while computer memory usage by legitimate applications usually varies.

The third indicator chosen was the amount of GPU memory used by application. Our mining software significantly relied on GPU memory, loading almost all available GPU memory at all time. While legitimate GPU-intensive applications also used GPU memory intensively, their load was somewhat varied with time, so an average GPU memory consumption was lower than 79%

Final indicator was the average quadratic deviation of graphic processor utilization share.

Analysis of the experimental data revealed that the average quadratic deviation of GPU consumption by mining processes remained quite low, below 3.1, whereas legitimate GPU-intensive applications exhibit more dynamic graphic processor utilization with a quadratic deviation no less that 4.7. Major exception to this was FurMark, the GPU benchmark, which exhibited similar GPU usage and GPU usage quadratic deviation to the real GPU miners. However, in our test environment it was still possible to distinguish benchmark with cryptojackers due to the memory metrics.

Additionally, no GPU cryptojacker consumed less than 96% of GPU memory and less than 85% of graphic processor time.

*C. Algorithm overview*

Implementing experimental data described above, we we have designed a decision tree algorithm, shown on Fig. 1. It was able to successfully detect all 3 mining processes mentioned in section A, while producing zero false positives on legitimate GPU-intensive applications.

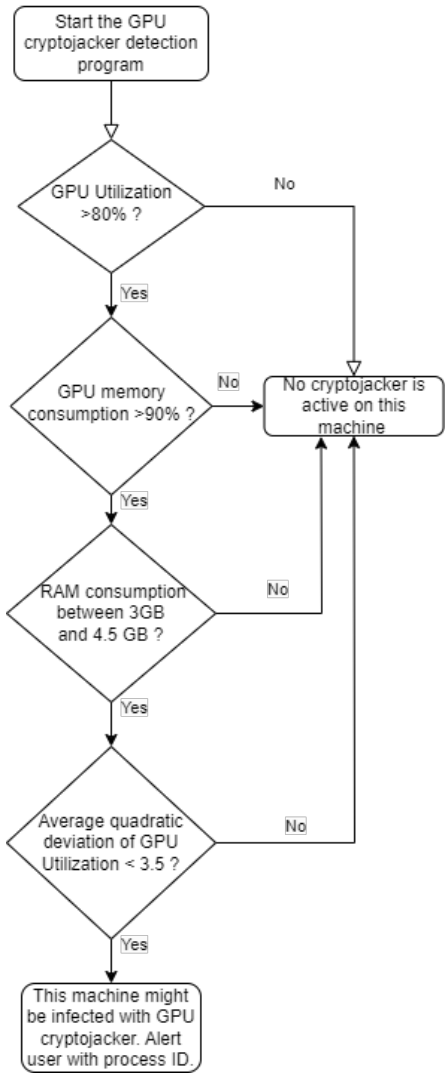

Fig. 1. GPU Cryptojacking malware detection algorithm.

Based on experimental data and algorithm on Fig. 1, we've developed a prototype GPU cryptojacking detection program. When this program detects suspicious activity on a computer, that might relate to GPU cryptojacking, it alerts the user with process ID. And then the decision on whether to terminate such process using standard Windows tools lies with user.

## IV. Results and Conclusion

To validate our detection algorithm, we've tested our GPU cryptojacker detection program on 10 more XMRig configurations, found on cybersecurity websites, which are supposed to emulate real cryptojacking activity as well as 5 more legitimate applications: Heaven Benchmark, PassMark, Maya, Baldur's Gate 3 and a Chrome browser tab running Youtube video.

Legitimate applications didn't trigger our detection program with the exception of PassMark, which performed similarly to the GPU miners both on GPU metrics and on memory metrics.

Out of 10 GPU miner configurations, our detection program was able to successfully identify 8. The two undetected configurations used stealth technique – continuously varying GPU load between 60 and 90 percent, thus evading detection by GPU metrics. Overall success rating for our detection program can be seen in table 1. Still, for a prototype program without reliance on machine learning techniques such results are quite promising.

TABLE I. Results for Detection Program

| Samples set | Number of samples | Detections |
|---|---|---|
| Test set | 3 | 3 |
| Validation set | 10 | 8 |
| Legitimate applications | 5 | 1 |

The GPU cryptojacking is an understudied threat in blockchain ecosystem which should be investigated to the level of CPU cryptojacking and ransomware. While this study is obviously limited in scope and test samples, in the future works we would like to enhance our algorithm and approbate it against live GPU cryptojacking samples on a special test stand.

Another aspect we'll be striving to improve is detection of GPU cryptojackers which use stealth technique, such as varying GPU load. Currently we think it would be possible to fine-tune GPU performance metrics in order to successfully identify GPU cryptojackers with varying GPU load, however live GPU cryptojackers might implement some other stealth technique forms, thus, requiring further studies.

Final problem we're going to look in the future is GPU cryptojacking detection in parallel to the presence of other GPU-heavy application, such as videogame, 3D modelling software or an AI training process. Our current findings don't make it clear if such detection is possible.